\documentclass{appolb}
\usepackage{epsfig}

\newcommand{\be}{\begin{equation}}
\newcommand{\ee}{\end{equation}}
\newcommand{\ba}{\begin{eqnarray}}
\newcommand{\ea}{\end{eqnarray}}
\newcommand{\ban}{\begin{eqnarray*}}
\newcommand{\ean}{\end{eqnarray*}}

\begin{document}

\title{Unstable Quark-Gluon Plasma at LHC\thanks{Presented at Cracow 
Epiphany Conference on LHC Physics, Cracow, Poland, January 4-6, 2008}}

\author{Stanis\l aw Mr\' owczy\' nski 
\address{Institute of Physics, Jan Kochanowski University \\
ul.~\'Swi\c etokrzyska 15, PL - 25-406 Kielce, Poland \\
and Andrzej So\l tan Institute for Nuclear Studies \\
ul.~Ho\.za 69, PL - 00-681 Warsaw, Poland}}

\date{2-nd May 2008}

\maketitle

\begin{abstract}

Coupling of the quark-gluon plasma from the early stage of heavy-ion 
collisions is argued to be significantly weaker at LHC than at RHIC. 
For this reason, the role of instabilities - the pre-equilibrium plasma 
is unstable with respect to chromomagnetic modes - will be enhanced. The 
instabilities isotropize the system and speed up the process of equilibration. 
A possibility to observe direct signals of the instabilities is considered.

\end{abstract}


\section{Introduction}

A prospect of heavy-ion program at Large Hadron Collider (LHC) to be
initiated at CERN soon, makes one to wonder about new features of 
nucleus-nucleus interactions when the collision energy is increased 
nearly 30 times when compared to the highest accessible energy at 
Relativistic Heavy Ion Collider (RHIC) at BNL. The problem actually
attracted a lot of attention and numerous predictions were formulated 
\cite{Abreu:2007kv}. In my lecture, however, I would like to focus on 
one specific aspect of heavy-ion collisions - the dynamical role of
chromomagnetic Weibel instabilities which, I expect, will be strongly 
enhanced at LHC. 

A successful experimental program at RHIC provided a convincing
evidence that the quark-gluon plasma (QGP), which is produced at the 
early stage of relativistic heavy-ion collisions, equilibrates fast - 
presumably within the time interval as short as 1 fm/$c$ - and 
later on it behaves as a nearly ideal fluid \cite{Heinz:2005ja}. Both 
features can be easily explained assuming that QGP is strongly coupled 
\cite{Shuryak:2004kh}. However, it is quite probable that the plasma 
at the pre-equilibrium stage of the collision is rather weakly coupled 
due to the asymptotic freedom regime achieved at a very high energy 
density. Actually, the very recent analysis \cite{Qin:2007rn} of the 
RHIC data on jet quenching has shown that even the equilibrium
QGP is not so strongly coupled, as the inferred coupling constant
$\alpha = 0.3$. And it has been argued that the weakly coupled plasma can 
actually equilibrate fast, for a review see \cite{Mrowczynski:2005ki}, 
and it can behave as a fluid of low viscosity \cite{Asakawa:2006jn}. 
This is possible because the pre-equilibrium plasma is unstable with 
respect to the chromomagnetic plasma modes due to the anisotropy of 
momentum distribution. The instabilities isotropize the system and 
efficiently speed up the equilibration process \cite{Mrowczynski:2005ki}. 
The spontaneously generated magnetic fields can be responsible for the 
so-called anomalous viscosity which make the plasma behave as a nearly 
ideal fluid \cite{Asakawa:2006jn}. However, the instabilities are 
operative if the plasma is weakly coupled. In the following sections 
I argue that conditions for the instabilities are much more favorable 
at LHC than at RHIC and I speculate about possible direct signals of 
the unstable modes.

\section{Coupling of QGP at LHC}

I argue here that, due to the increase of energy density, the coupling 
of QGP at the pre-equilibrium stage of relativistic-heavy-ion 
collisions is significantly weaker at LHC than at RHIC. 

The (thermal) energy density and temperature of the plasma at the 
moment when it reaches local thermodynamic equilibrium are estimated 
as $\epsilon^{\rm RHIC}_T \approx 30 \; {\rm GeV/fm^3}$ and 
$T^{\rm RHIC} \approx 350 \; {\rm MeV}$ (the lower index $T$ stands
for `thermal') at the highest RHIC energy ($\sqrt{s}$ = 200 GeV per 
N-N collsion) \cite{Heinz:2005ja,Broniowski:2008vp}. The analogous 
quantities for LHC ($\sqrt{s}$ = 5500 GeV per N-N collision) are 
expected to be $\epsilon^{\rm LHC}_T \approx 130 \; {\rm GeV/fm^3}$ 
and $T^{\rm RHIC} \approx 500 \; {\rm MeV}$
\cite{Broniowski:2008vp,Bluhm:2007sf}. The increase of the 
temperature is not very big but noticeable and it influences
the plasma coupling. In the ideal gas of massless particles 
$\epsilon_T = 3p$ where $p$ is the gas pressure. Therefore, the 
dimensionless quantity $(\epsilon_T - 3p)/T^4$ is often treated as 
a measure of the interaction strength in the plasma. The QCD lattice 
calculations show that $(\epsilon_T - 3p)/T^4 \approx 3$ at $T$ = 350 MeV 
and it is reduced to about unity at $T$ = 500 MeV \cite{Karsch:2007dp}. 
Thus, the early stage plasma is closer to the non-interacting gas at LHC 
than at RHIC.

However, we are mostly interested in the pre-equilibrium plasma and we would 
like to get an estimate of the coupling constant $\alpha \equiv g^2/4\pi$ 
which at RHIC energies is usually chosen to be $\alpha^{\rm RHIC} \approx 0.3$ 
\cite{Wang:1996yf}. As already mentioned, the very recent analysis of the RHIC 
data on jet quenching supports correctness of this choice \cite{Qin:2007rn}. 
Let us first get an idea about the energy density in the center-of-mass frame 
just after the collision. In the central collisions of nuclei of mass number 
$A$, we estimate it as 
\be
\epsilon_0 = \frac{k A \sqrt{s}}{\pi R^2 l} \;,
\ee
where $k$ is the inelasticity - the fraction of initial energy, which 
goes to particle production, $R$ is the radius of colliding nuclei and 
$l$ is the length of the cylinder where the energy is released. Assuming that 
$k=0.5$ independently of energy \cite{Navarra:2003am} and taking $A=200$, 
$R = 7$ fm and $l = 1$ fm, one obtains  $\epsilon^{\rm RHIC}_0 \approx 130 
\; {\rm GeV/fm^3}$ for $\sqrt{s}$ = 200 GeV and and $\epsilon^{\rm LHC}_0 
\approx 3600 \; {\rm GeV/fm^3}$ for $\sqrt{s}$ = 5500 GeV. One wonders why 
$\epsilon_T$ is so much smaller than $\epsilon_0$ (4 times for RHIC and 28 
for LHC). The system's expansion during the pre-equilibrium phase is partially 
responsible for the decrease but it is far more important that the thermal 
energy density does {\em not} include the energy related to a collective 
motion which is very large. 

The evolution of $\alpha$ with energy density can be roughly estimated, 
using the celebrated formula of running coupling constant  
\be
\label{alpha}
\alpha (Q^2) = \frac{12\pi}{(33-2N_f)\ln (Q^2/\Lambda_{\rm QCD}^2)} \;,
\ee
where $Q$ is the characteristic momentum transfer, $\Lambda_{\rm QCD}=200$
MeV is the QCD scale parameter and $N_f = 3$ is the number of flavours.
Referring to the dimensional argument, I replace $Q^2$ by 
$a \sqrt{\epsilon_0}$ with the parameter $a = 4.12$, which is chosen 
in such a way that Eq.~(\ref{alpha}) with $\epsilon_0 = \epsilon^{\rm RHIC}_0$
gives $\alpha^{\rm RHIC} = 0.3$. Then, I obtain $\alpha^{\rm LHC} = 0.16$ for 
$\epsilon_0 = \epsilon^{\rm LHC}_0$. The reduction appears to be quite 
significant but the actual numerical value of $\alpha$ is, obviously, not
very reliable. However, I can safely conclude this section that the 
lattice calculations and the asymptotic freedom argument both suggest 
noticeably weaker coupling of the early stage QGP at LHC than at RHIC.

\section{Instabilities at LHC}

Occurrence of Weibel instabilities requires an anisotropy of parton 
momentum distribution. The condition is trivially fulfilled, as initially
the parton momentum is strongly elongated along the beam - its
shape is prolate - and in the course of system's expansion it becomes 
locally squeezed along the beam - its shape is oblate. And the Weibel
modes are present in both configurations. 

The anisotropy can be quantified by means of the parameter 
$2 \langle p_L^2 \rangle/\langle p_T^2 \rangle$, where $p_L$ and 
$p_T$ are parton longitudinal and transverse momenta and 
$\langle \dots \rangle$ denotes inclusive averaging. As discussed 
in \cite{Jas:2007rw}, the parameter is well approximated by the formula
\be
\frac{2 \langle p_L^2 \rangle}{\langle p_T^2 \rangle}= e^{2\sigma^2_y} - 1,
\ee
where partons are assumed to be massless and $\sigma_y$ is the width of 
Gaussian rapidity distribution. To obtain $\sigma_y$ at LHC, I use the 
Landau model parameterization \cite{Carruthers:1973ws}
\be
\label{y-width}
\sigma_y^2 = \ln \big(\sqrt{s}/2m_p\big) \;,
\ee
where $m_p$ is the proton mass. Eq.~(\ref{y-width}) gives 
$\sigma_y=2.2$ at the top RHIC energy which agrees well with the 
pion data \cite{Bearden:2004yx}. For $\sqrt{s} = 5500$ GeV, the 
formula (\ref{y-width}) predicts $\sigma_y=2.8$ which in turn gives 
$2 \langle p_L^2 \rangle/\langle p_T^2 \rangle = 7 \cdot 10^6$
(at the top RHIC energy 
$2 \langle p_L^2 \rangle/\langle p_T^2 \rangle = 2 \cdot 10^4$).
So, initially there will be a huge anisotropy which will locally decay 
in the course of system's expansion. As shown in \cite{Jas:2007rw},
it takes about 5 fm/$c$ to make the local distribution oblate
($2 \langle p_L^2 \rangle/\langle p_T^2 \rangle < 1$), if the 
system evolves solely due to the free streaming. An actual
evolution of the momentum distribution will be faster, as not only
the free streaming but interactions in the parton system will tend 
to reduce the momentum anisotropy, but one expects that during the 
first, say, 1-2 fm/$c$ the unstable modes will be operative
if their growth rates are sufficiently large. As I explain below,
this requires a weak coupling and high energy density of the system. 

If QGP is really weakly coupled, the characteristic inverse time of 
processes driven by inter-parton collisions is 
$t_{\rm hard}^{-1} \sim g^4 {\ln}(1/g)p$ or 
$t_{\rm soft}^{-1} \sim g^2 {\ln}(1/g)p$, depending whether the 
momentum transfer in a collision is of order $p$ or $gp$ with $p$ being 
a typical parton momentum which in the equilibrium plasma can be
identified with the temperature $T$ \cite{Arnold:1998cy}. The characteristic 
inverse time of mean-field collective phenomena, in particular the growth 
rate of instabilities is of order $gp$ \cite{Mrowczynski:2005ki}. 
Therefore, there is a good separation of the time scales provided 
$g^2 \ll 1$. Then, the instabilities are much faster than the inter-parton 
collisions which are responsible for dissipative processes. The collisions 
slow down the growth of the unstable modes and there is an upper limit on 
the collisional frequency beyond which no instabilities exist 
\cite{Schenke:2006xu}. Therefore, the Weibel instabilities require
a sufficiently weak coupling of the plasma.

While the weak coupling guarantees that the instabilities are faster
than the collisions, the instabilities should be also much faster than
the characteristic time of the system's temporal evolution. Then, the
spontaneously generated chromomagnetic fields will reach large values 
as a result of many e-foldings of their amplitudes. As already 
mentioned, the growth rate of the instabilities $\gamma$ is of order $gp$ 
\cite{Mrowczynski:2005ki}. If $p$ is identified with $\epsilon_0^{1/4}$,
where $\epsilon_0$ is the initial energy density discussed earlier,
$p^{\rm LHC} = 2.3$ GeV (for top RHIC energy $p^{\rm RHIC} = 1.0$ GeV). 
As the coupling constant $g$ is actually of order of unity, the 
instabilities are presumably faster than the characteristic time of 
the system's temporal evolution ($\gamma^{-1} \sim 0.1$ fm/$c$), but 
it is unclear whether the instabilities are fast enough to avoid 
a strong damping caused by the inter-parton collisions. 

Although my estimates do not allow to draw a firm conclusion that 
the Weibel instabilities will indeed play an important role in the 
pre-equilibrium QGP at LHC, my estimates clearly show that the 
conditions will be much more preferable than those at RHIC. And if 
the instabilities already exist at RHIC, as the fast thermalization 
suggests, they should play a prominent role at LHC. So, let me discus 
how the Weibel instabilities can manifest themselves.

\section{Fast Equilibration}
\label{sec-fast-eq}

The instabilities speed up the process of equilibration as they 
effectively isotropize the momentum distribution. It should be 
stressed here that inter-parton collisions are not very effective 
in changing parton's momenta, because the one-gluon-exchange cross section, 
as the Rutherford one, is strongly peaked at small momentum transfers. 
The characteristic time of collisional izotropization coincide with 
the earlier introduced $t_{\rm hard}$ and it is too long in the weakly 
coupled plasma to comply with the fast equilibration. The radiative 
parton collisions are more effective in redistributing parton's momenta 
but the processes are suppressed by an extra power of $\alpha$. The 
magnetic fields associated with the unstable modes do the job really 
fast.

To explain the mechanism of isotropization I assume that the momentum 
distribution is strongly elongated along the beam ($z$) direction. 
The colour currents, which initiate the Weibel instability as a random 
fluctuation and then grow when the instability develops, flow in the 
$z$ direction. The wave vector of the fastest unstable mode lies in the 
$x\!-\!y$ plane. I assume that it points in the $x$ direction. Then, the 
magnetic field generated by the currents is oriented in the $y$ direction 
and the Lorentz force, which acts on partons flying along the $z$ axis, 
pushes them in the $x$ direction where there is a momentum deficit. Having 
a superposition of many unstable modes with their vectors in the 
$x\!-\!y$ plane, the instabilities produce approximately axially 
symmetric transverse momentum. 

The system isotropizes not only due to the effect of the Lorentz
force but also due to the momentum carried by the growing field. 
When the magnetic and electric fields are oriented along the
$y$ and $z$ axes, respectively, the Poynting vector points in
the direction $x$ that is along the wave vector. Thus, the momentum 
carried by the fields is oriented in the direction of the momentum 
deficit. The numerical simulations \cite{Dumitru:2005gp,Berges:2007re}, 
which, however, were performed for the oblate not prolate momentum 
distribution, indeed show that growth of instabilities is accompanied 
by the system's fast isotropization. 

The isotropization should not be confused with the equilibration.
Obviously, the instabilities cannot equilibrate the system but
once the instabilities have redistributed parton's momenta, soft 
parton-parton collisions, which are much more frequent than the 
hard ones, complete the processes of equilibration.

\section{Signals of instabilities}

Since the Weibel instability is a phenomenon, which occurs at the
pre-equilibrium phase of QGP, later temporal evolution of the plasma,
including equilibration and hadronization, conceals its presence 
and it is difficult to point a direct signal of the instabilities 
which can be observed in the final state of heavy-ion collisions. 

Recently it has been argued \cite{Majumder:2006wi} that the 
experimentally observed longitudinal broadening of jets, which are 
quenched in the plasma, can be attributed to the interaction of jet
particles with colour fields generated by the unstable modes. I would 
like to briefly discuss two related to each another possible signals 
of the instabilities. 

\subsection{Elliptic flow fluctuations}

The so-called elliptic flow, which is caused by an initially asymmetric
shape of the interaction zone of colliding nuclei, is sensitive to the 
collision early stage. The phenomenon is successfully described by the hydrodynamic model, see e.g. \cite{Heinz:2005ja}, which, in principle,
requires that the system under study is in a local thermodynamical 
equilibrium. However, an approximate hydrodynamic behaviour occurs, 
as argued in \cite{Arnold:2004ti}, when the momentum distribution of 
liquid components is merely isotropic in the local rest frame. The point 
is that the structure of the ideal fluid energy-momentum tensor {\it i.e.} 
$T^{\mu \nu} = (\varepsilon + p) \, u^{\mu} u^{\nu} -p \, g^{\mu \nu}$, 
where $\varepsilon$, $p$ and $u^{\mu}$ is the energy density, pressure 
and hydrodynamic velocity, respectively, holds for an arbitrary, though 
isotropic momentum distribution. $\varepsilon$ and $p$ are then not the 
energy density and pressure but the moments of the distribution function 
which are equal the energy density and pressure in the equilibrium limit. 
Since the tensor $T^{\mu \nu}$ obeys the continuity equation 
$\partial_\mu T^{\mu \nu} =0$, one gets an analogue of the Euler 
equation. However, due to the lack of thermodynamic equilibrium
there is no entropy conservation.

Usually, non-equilibrium fluctuations are significantly bigger
than the equilibrium fluctuations of the same quantity. Therefore, 
sizeable fluctuations of the elliptic flow due to the pre-equilibrium 
stage of quasi-hydrodynamic evolution were predicted 
\cite{Mrowczynski:2002bw,Mrowczynski:2005gw}. Large fluctuations 
of the elliptic flow have been indeed observed at RHIC 
\cite{Alver:2007rm,Sorensen:2006nw} but the effect is now commonly
understood as a result of fluctuations of the eccentricity of the
interaction zone as suggested in \cite{Miller:2003kd}. Thus, it is
assumed that the whole effect of the observed elliptic flow fluctuations 
comes from the eccentricity fluctuations while the hydrodynamic evolution 
is assumed to be fully deterministic and thus, it does not contribute to 
the elliptic flow fluctuations\footnote{Very recently the observation of 
large elliptic flow fluctuations has been retracted by STAR collaboration 
\cite{Sorensen:2006nw-2}. It is claimed now that the previously given 
magnitude of the fluctuations should be taken only as an upper limit 
due to the difficulties to disentangle the elliptic flow fluctuations 
and the contributions which are not correlated with the reaction plane. 
Since the effects of instabilities are not associated with the reaction 
plane, the retraction \cite{Sorensen:2006nw-2} does not much influence 
the considerations presented here.}.

Although the calculations of the eccentricity fluctuations reproduce 
well the experimentally observed elliptic flow fluctuations, see e.g. \cite{Broniowski:2007ft}, the eccentricity fluctuations seem to me 
significantly overestimated. The nucleons of colliding nuclei are 
treated as essentially independent from each other, and consequently 
a smooth shell structure of a nucleus is ignored. It is even more
important that a significant contribution to the eccentricity fluctuations 
comes from collisions of nucleons from a nucleus periphery. Transverse
positions of these collisions are usually well separated from the  
positions of interactions of other nucleons \cite{Broniowski-private}.
In my opinion, partons produced in these isolated nucleon-nucleon 
collisions do not participate in the hydrodynamic evolution of the 
system and do not contribute to the elliptic flow. If the isolated 
collisions are excluded, the eccentricity fluctuations are reduced, 
and an extra source of fluctuations is needed to explain the data. 
The extra fluctuations come, I suppose, from the quasi-equilibrium 
hydrodynamic evolution. 

\subsection{Azimuthal fluctuations}

As argued in the previous subsection, the instability driven
equilibration leads to significant elliptic flow fluctuations 
because the quasi-hydrodynamic evolution starts when the
system is merely isotropic but not fully equilibrated. However,
the instabilities can also be directly responsible for flow-like 
effects. In the prolate momentum configuration, which is 
characteristic for the pre-equilibrium stage of the quark-gluon
plasma \cite{Jas:2007rw}, the wave vector of the fastest unstable 
modes is randomly oriented in the transverse plane. As explained
in Sec.~\ref{sec-fast-eq}, the momentum is transported along 
the wave vector. Therefore, I expect a collective radial flow 
which exhibits strong azimuthal fluctuations \cite{Mrowczynski:2005gw}. 
In contrast to the elliptic flow, the transverse flow caused by 
the instabilities is {\em not} correlated with the reaction plane.
In particular, the flow should occur in exactly central collisions 
when the elliptic flow is absent for the symmetry reasons.  
 
\section{Conclusions}

The Weibel instabilities seem to be present in relativistic 
heavy-ion collisions at RHIC and the estimates show that the 
conditions for the instabilities will be much more preferable at 
LHC than at RHIC. Therefore, I expect that the role of 
instabilities will be strongly enhanced.

The instabilities provide a plausible mechanism responsible for 
a surprisingly short equilibration time and the fast isotropization 
is a distinctive feature of the mechanism.  Since the Weibel 
instabilities occur at the pre-equilibrium phase of QGP, later 
temporal evolution of the plasma conceals their presence and it is 
difficult to indicate direct signals. However, careful studies of the 
evolution of jets, of the azimuthal fluctuations of radial flow and of 
the elliptic flow fluctuations will hopefully provide an evidence that
the pre-equilibrium QGP is indeed unstable.

\vspace{1cm}

I am grateful to Uli Heinz and Constantin Loizides for helpful 
correspondence and to Art Poskanzer and Xin-Nian Wang for 
fruitful discussions. 


\end{document}